\documentclass[aps,prl,letterpaper,twocolumn,showpacs,superscriptaddress,floatfix,longbibliography]{revtex4-1}

\usepackage{amsmath}
\usepackage{amssymb}

\usepackage{amsmath,amssymb}
\usepackage[usenames]{color}
\usepackage{mathbbol}
\usepackage{amssymb}
\usepackage{grffile}
\usepackage[pdftex]{graphicx}
\usepackage{amsmath, amstext, amssymb, amsfonts, amsxtra}
\usepackage{textcomp}
\usepackage{xspace}
\usepackage[colorlinks]{hyperref}

\usepackage[english]{babel}
\usepackage{amsmath}
\usepackage{color}
\usepackage{epsfig}
\usepackage{graphicx}
\usepackage{bm}
\usepackage{times,amsmath,amssymb}
\usepackage{subfigure}
\usepackage{amsfonts}
\usepackage{amssymb}
\usepackage{color}

\usepackage{mathtools}
\usepackage{empheq}
\usepackage{pifont}

\usepackage[dvipsnames]{xcolor}
\usepackage{color}

\DeclareSymbolFontAlphabet{\amsmathbb}{AMSb}

\definecolor{myred}{RGB}{168,5,14}

\definecolor{myblue}{RGB}{14,5,168}

\global\long\def\bra#1{\langle #1 \vert}
\global\long\def\ket#1{\vert #1\rangle }
\global\long\def\braket#1#2{\left\langle #1|#2\right\rangle }

\global\long\def\al{\alpha}
\global\long\def\be{\beta}

\global\long\def\Ga{\Gamma}

\global\long\def\vfi{\varphi}

\global\long\def\bege{\begin{equation}}
\global\long\def\ende{\end{equation}}

\global\long\def\begal{\begin{align}}
\global\long\def\endal{\end{align}}

\newcommand{\Hop}{\hat{H}}

\newcommand{\Dop}{\mathcal{D}}

\newcommand{\im}{{\rm i}}

\newcommand{\spare}[1]{\left[#1 \right]}

\newcommand{\bonn}{HISKP, University of Bonn, Nussallee 14-16, 53115 Bonn, Germany}
\newcommand{\bonnpi}{Physikalisches Institut, University of Bonn, Nussallee 12, 53115 Bonn, Germany}

\begin{document}

\title{Transition between dissipatively stabilized helical states}

\author{Simon Essink}
\affiliation{\bonnpi}
\author{Stefan Wolff}
\affiliation{\bonnpi}
\author{Gunter M. Sch\"utz}
\affiliation{Institute of Complex Systems II, Forschungszentrum J\"{u}lich - 52425 J\"{u}lich, Germany}
\author{Corinna Kollath}
\affiliation{\bonnpi}
\author{Vladislav Popkov}
\affiliation{\bonn}
\affiliation{Department of Physics,  University of Ljubljana,
Jadranska 19, SI-1000 Ljubljana, Slovenia}
\affiliation{Department of Physics, Bergische Universit\"at Wuppertal
Gaussstr.20, 42097 Wuppertal
Germany}

\begin{abstract}
  We analyze a $XXZ$ spin-$1/2$ chain which is driven dissipatively at its boundaries. The dissipative driving is modelled by Lindblad jump operators which only act on both boundary spins. In the limit of large dissipation, we find that the boundary spins are pinned to a certain value and at special values of the interaction anisotropy, the steady states are formed by a rank-2 mixture of helical states with opposite winding numbers. Contrarily to previous stabilization of topological states, these helical states are not protected by a gap in the spectrum of the Lindbladian. By changing the anisotropy, the transition between these steady states takes place via mixed states of higher rank. In particular, crossing the value of zero anisotropy a totally mixed state is found as the steady state. The transition between the different winding numbers via mixed states can be seen in the light of the transitions between different topological states in dissipatively driven systems.  The results are obtained developing a perturbation theory in the inverse dissipative coupling strength and using the numerical exact diagonalization and matrix product state methods.
\end{abstract}

\date{\today}
\maketitle


Over decades, dissipation has been considered as a destructive influence which destroys coherence properties of quantum systems. Recently, this point of view has been revised, since tailored environments have been employed in order to dissipatively drive a quantum many body system into a
desired steady state, the so-called attractor state \cite{MuellerZoller2012}. Even if an external perturbation is applied over a certain time window, the system flows back to the attractor state afterwards. Examples of many-body states that can be reached via an attractor dynamics of a tailored environment are Bose-Einstein condensates \cite{DiehlZoller2008}, number squeezed states \cite{CaballarWatanabe2014}, Tonks-like states \cite{SyassenDuerr2008}, superconducting states \cite{DiehlZoller2010,SheikhanKollath2019}, and, more recently, topologically interesting states \cite{BardynImamoglu2012b, BardynDiehl2013,SheikhanKollath2016,KollathBrennecke2016,IeminiMazza2016}. These comprise Chern insulators \cite{BudichDiehl2015} and the Hofstadter model of atoms in an optical cavity \cite{SheikhanKollath2016}.

Topological states are characterized by the existence of invariants which  can only change in steps by a global action on the system. A paradigmatic example is the use of the stepwise change of the electrical resistance in the quantum Hall effect in topological insulators which is employed for the definition of the standard for the electrical resistance \cite{Ezawa2013}.
The classification of topological properties in non-interacting closed systems has attracted considerable attention \cite{HasanKane2010,XiaoNiu2010,HasanNeupane2015}. In contrast, topological properties in interacting or open quantum systems are much less well understood, despite intensified efforts during the last few years.

In open non-interacting systems \cite{BardynDiehl2013} two important ingredients were identified for reaching stable topologically non-trivial states. The first one is the existence of a dissipative gap, i.e. a gap in the spectrum of the Lindbladian above the steady state.
 The second one can only be introduced in non-interacting systems and is the so-called purity gap. This gap measures the purity of the most strongly mixed mode of the bulk.

Here we go far beyond current studies and show how the intriguing interplay of interactions and a tailored dissipative coupling can give access to novel topologically interesting properties. To do this, we study by exact analytical and numerical methods the paradigmatic spin-$1/2$ $XXZ$-quantum spin chain with dissipative boundaries. Previous work has uncovered far-from-equilibrium steady states of helical nature \cite{PopkovSchmidt2017,PosskeThorwart2019} with remarkable transport properties \cite{PopkovSchuetz2017}. In this work, we focus on the case that the dissipative jump operators at the boundary sites of the chain are identical which leads to an additional reflection symmetry. We find that in the limit of large dissipation the space reflection symmetry of the system leads to the situation that at certain discrete values of the anisotropy parameter rank-2 steady states -- formed by helical states with opposite winding numbers -- are dissipatively generated. These winding numbers have integer values and therefore, similar to topological invariants, can only change their values in integer steps.

 The helical steady states are not protected by a finite gap which is in contrast to topological states in open systems found previously \cite{BardynDiehl2013}. As one varies the interaction strength a transition between two helical states occurs, which takes place via higher rank mixtures of states to which several different winding numbers contribute. When the anisotropy changes sign, the steady state transits even via a completely mixed state.

%
%

We describe the $XXZ$ chain with density operator $\rho$ by the Lindblad master equation
\begin{align}
\frac{d\rho}{dt} = -\frac{\im}{\hbar} \spare{\Hop,\rho} + \Dop (\rho) \label{eq:master}.
\end{align}
Below we shall set $\hbar=1$.
The first term on the right-hand side describes the unitary evolution due to the $XXZ$-Hamiltonian
\begin{align}
\Hop=J\sum_{j=1}^{N-1} \left[S^x_j S^x_{j+1} +  S^y_j S^y_{j+1} + \Delta (S^z_j S^z_{j+1}-\frac{1}{4}I) \right].
\end{align}
Here $S^\alpha_j=\sigma^\alpha_j/2$ are the spin-$1/2$ operators and $\sigma_j^\alpha$ the Pauli matrices acting on site $j$. The parameter $\Delta$ is the anisotropy which determines the quantum phases that appear in an isolated system. The identity $I$ is added for convenience. For $|\Delta| \le 1$ the ground state of the $XXZ$-Hamiltonian is a gapless Tomonaga-Luttinger liquid. For values $|\Delta|>1$ a gapped phase occurs which corresponds to a ferromagnetic or antiferromagnetic ground state, respectively.
$N$ is the number of sites and we assume in the following for convenience $N$ to be an even number.

The second term describes the dissipative coupling to the environment in Lindblad form $\mathcal{D} [\rho]=\mathcal{D}_1 [\rho]+
\mathcal{D}_N [\rho]$, where
\begin{equation}
 \mathcal{D}_j [\rho] =\Gamma \left( L_j \rho L_j^\dagger - \frac{1}{2} L_j^\dagger L_j
    \rho - \frac{1}{2}\rho L_j^\dagger L_j\right).
  \label{DefLMEDissipator}
\end{equation}
Here $\Gamma$ is the effective dissipation strength, and $L_j$ are the jump operators which act only at the boundary sites $j=1$ and $j=N$ and target the density matrix belonging to the eigenstate $\ket{\uparrow_x}$  of the spin operator in $x$ direction defined by $ \sigma^x\ket{\uparrow_x}=\ket{\uparrow_x}$. Explicitly,
$L_1= S^y_1+i S^z_1 $ and $L_N= S^y_N+iS^z_N $. We can show that in this situation a unique steady state exists \cite{Prosen2012}.

In the Zeno limit of large dissipative coupling $\Gamma\to \infty$, the boundary spins to lowest order are pinned in the steady state to the states defined by $\mathcal{D}_{1,N}[\ket{\uparrow_x}_{1,N}\bra{\uparrow_x}_{1,N}]=0$.
The dissipation free subspace of the system is thus the whole Hilbert space spanned by the bulk spins and fixed
 boundary spins $1,N$  which are collinear and oriented in the positive $x$ direction, i.e.~$\ket{\uparrow_x}$.

Previous studies \cite{PopkovPresilla2016,PopkovSchmidt2017,PopkovSchuetz2017} have found that for many choices of the boundary dissipation a fine-tuning of the anisotropy $\Delta^*_m=\cos (\vfi_m+\delta \vfi/(N-1)) $ with the angle $\vfi_m=(2 \pi m)/(N-1)$ with $m= -N/2 \ldots N/2$,
generates a pure steady state which is a spin-helix state
\begin{align}
 \ket{m}  &= \frac{1}{\sqrt{2^N}} \bigotimes_{j=1}^{N} \left(
    \begin{array}{c}
        \mathrm{e}^{-\frac{i}{2}(j-1) (\varphi_m+\delta \vfi)}
      \\
      \mathrm{e}^{ \frac{i}{2}(j-1) (\varphi_m+\delta \vfi)}
    \end{array}
  \right).
  \label{SHS}
\end{align}
where $\delta \vfi$ is a twist angle between the targeted boundary polarizations.
Here the state on each site is represented in the basis chosen along $z$-direction and the spin precesses in the $XY$-plane around the $z$-axis.
However, this steady state will become  unstable if the spin states targeted at the boundaries
become collinear $\delta \vfi=0$ and $m\ne 0$, as is the case in the chosen situation. We found similar results for $\delta \vfi=\pi$.

For the situation, where the spins are locked to the dissipation free subspace, the system can be viewed as a spin chain on a ring, where the site $1$ and $N$ are glued to the same site. Within this configuration, important quantities are the winding numbers of the spin along the ring. They can be determined by the discrete Fourier transform
\begin{align}
 w_m = \frac{2}{N-1} \sum_{j=1}^{N-1} \langle S^+_j\rangle \mathrm{e}^{-i \varphi_m(j-1)}.
 \label{fourier_coefficients}
\end{align}
where $m = -(N-2)/2 \dots, (N-2)/2$ denotes the winding number around the $z$-axis and the amplitudes $w_m$ can be interpreted as the corresponding weights. Due to the symmetry of the considered system, the relation $w_m=w_{-m}$ holds.

We note that in a finite system states corresponding to different winding numbers can overlap. However, this overlap vanishes exponentially with system size. In the limit of infinite system size  the states corresponding to different winding numbers become orthogonal and the winding number corresponds to a topological invariant.

%
%

\begin{figure}
\includegraphics[width=0.9\columnwidth]{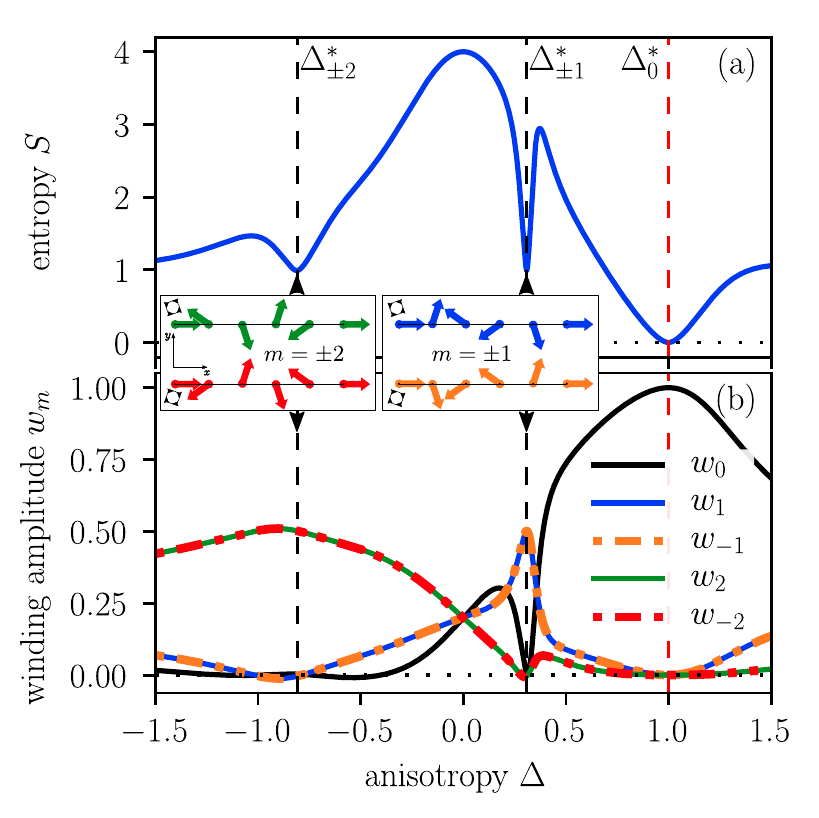}
\caption{(color online) (a) Von Neuman entropy and (b) Winding amplitudes versus the anisotropy $\Delta$ 
  for $N=6$ and $\Gamma=250J$ obtained by exact diagonalization. The black dashed vertical lines mark the special values of $\Delta^*$, where the rank-2 helical state is expected in the Zeno limit. The insets in (b) show the spin orientation of the two helical components of the state projected into the xy-plane at the special points.  The red dashed vertical line marks the value $\Delta=1$.
  \label{fig:diagram} }
\end{figure}

An intriguing behaviour can be seen in the von Neumann entropy
$ S = - \sum_i p_i \,\text{log}_2(p_i)$, where $p_i$ are the eigenvalues of the density matrix $\rho$. In Fig.~\ref{fig:diagram} we show the dependence of the von Neumann entropy on the anisotropy $\Delta$ for a small system ($N=6$) and a strong amplitude of the dissipative driving $\Gamma=250J$. For this small system, exact diagonalization is used to solve the quantum master equation (\ref{eq:master}). A drastic behaviour in the von Neumann entropy  can be seen at the values $\Delta^*_m=\cos \vfi_m$ with the angle $\vfi_m=2 \pi m/5$ with $m \in \{1,2\}$. At $\Delta^*_m$ two amplitudes $w_{\pm m}$ of the winding numbers become dominant,
 whereas the other values become negligible. This signals that a helical state of rank two with two opposite winding numbers arises as the steady state.

We will show that the steady state in the Zeno limit is of the form
\begin{align}
  \rho^{(0)} =
  \ket{\uparrow_x}_1\bra{\uparrow_x}_1 \otimes \left(
    b \ket{s}\bra{s} +
    (1-b) \ket{a}\bra{a} \right) \otimes
   \ket{\uparrow_x}_N\bra{\uparrow_x}_N,
  \label{eq:ResultNESSrank2}
\end{align}
with $\ket{s}$,
$\ket{a}$ being  orthogonal linear combinations of the spin-helix
states  $\ket{\pm m}|_{\textrm{bulk}}$, restricted to sites $2,\ldots ,N-1$, with opposite chiralities, $\ket{s}= A_s \left(\ket{m}+\ket{\! -\!m}\right)|_{\textrm{bulk}} $ and $\ket{a}= A_a (\ket{m}- \ket{\! -\!m} )|_{\textrm{bulk}}  $ with $b$, $A_s$ and $A_a$ weights.
Additional particularities occur at $\Delta=1$, where the entropy drops to zero, signaling a pure state which is a helical state corresponding to the winding number $m=0$, and at $\Delta=0$ where a totally mixed state appears.

\begin{figure}
\includegraphics[width=0.9\columnwidth]{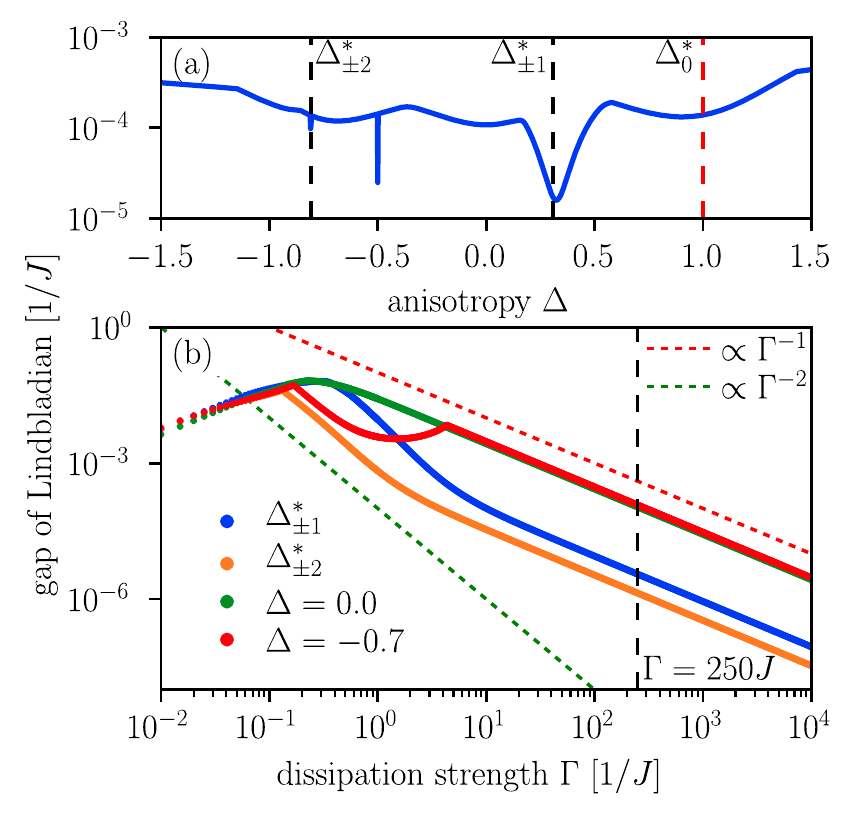}
\caption{(color online) (a) The dependence of the gap above the steady state in the Lindblad operator spectrum versus anisotropy $\Delta$ discretized in steps of 0.002 for $\Gamma=250J$ and $N=6$ calculated using exact diagonalization. (b) Dependence of the gap on the dissipation strength $\Gamma$ at different values of the anisotropy $\Delta$ for $N=6$. As a comparison the algebraic decay as $(J/\Gamma)^{\alpha}$ with $\alpha=1,2$ are plotted as dotted lines.
  \label{fig:gap}}
\end{figure}

This result demonstrates that steady states with different winding numbers can be reached by a fine tuning of the anisotropy. For finite dissipation strength $\Gamma$ and fine-tuned anisotropies, we find  numerically (not shown) that the steady state is approached as $tr(\rho^2(\Gamma)-tr((\rho^{(0)})^2 ) \propto (J/\Gamma)^2$, where $\rho(\Gamma)$ denotes the steady state at a finite value of $\Gamma$. The states at $\Delta^*$ seem not to be protected by a gap in the spectrum of the Lindbladian as can be seen from Fig.~\ref{fig:gap} where we show that the gap in the Zeno limit  closes as $1/\Gamma$.  This is in contrast to previous findings, where the topologically interesting states were protected by a gap \cite{BardynDiehl2013}. Further, the transition from one helical state to the other goes via the intermediate values of the anisotropy. In Fig.~\ref{fig:diagram}, this transition is performed via states which are composed of many different winding numbers and have a much larger von Neumann entropy. For the point which is close to $\Delta_{\pm 2}^*$, we have a relatively slow dependence, whereas the point corresponding to $\Delta_{\pm 1}^*$ has a very steep dependence.  Let us note, that the behaviour around the special points $\Delta^*$ steepens with increasing system length.

In order to verify that this is not just a particularity of the small system size, we used a purification implementation of the matrix product state (MPS) method for open quantum systems \cite{ZwolakVidal2004,VerstraeteCirac2004b,Schollwoeck2011} as described in Ref.~\cite{WolffKollath2019} to determine the steady states for larger systems. We have chosen to double the system size to $N=12$.

To obtain the steady state, we use the time-dependent MPS method based on a second order Suzuki-Trotter decomposition with time step $\Delta t$ to compute the long-time evolution of an arbitrary state which in this case is chosen to be the N\'eel state. To overcome the problem of slow relaxation during the attractor dynamics we employ a gradual time evolution procedure. As we are only interested in the steady state and the exact dynamics is irrelevant, we first apply an evolution in a fast-relaxing parameter regime to prepare the initial state $\rho^{\textrm{ini}}$ for the final evolution (see suppl.~for details).

\begin{figure}
\includegraphics[width=0.99\columnwidth]{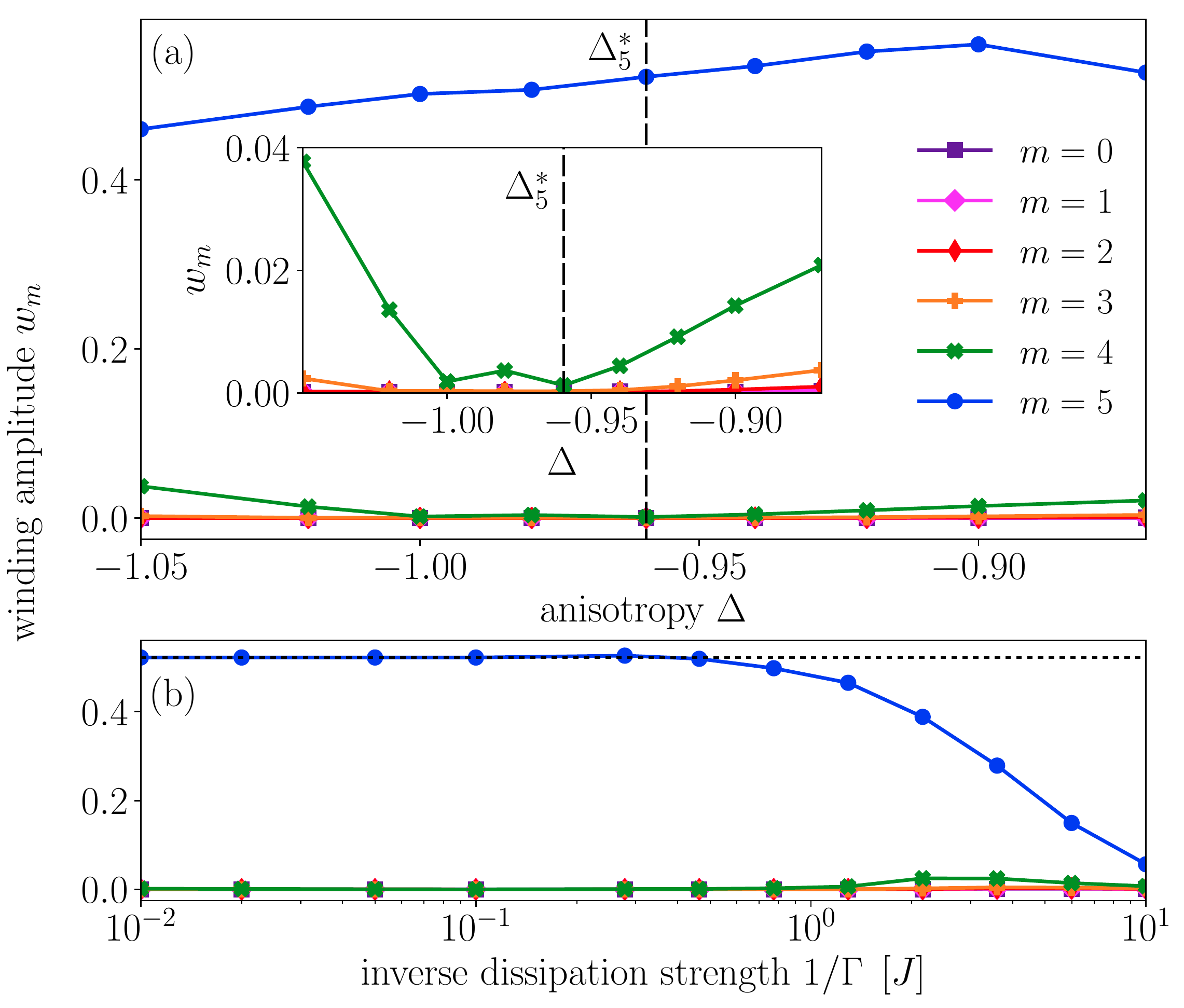}
\caption{(color online) (a) Winding amplitude $w_m$ versus anisotropy $\Delta$ around the value $\Delta^\ast_{\pm 5}$  for $\Gamma =50J$. (b) Winding amplitudes versus inverse dissipation strength $1/\Gamma$ at $\Delta^\ast_{\pm 5}$. The inset shows a zoom of small value regions and the results have been obtained for a system $N=12$, a maximal truncation weight $\epsilon=10^{-12}$ and a time step $\Delta t = 0.1/J$.
  \label{fig:L12} }
\end{figure}

This enables us to provide simulation results for different parameter ranges of the interaction anisotropy $\Delta$ and the dissipative coupling $\Gamma$.
 The simulation is based on an efficient compression scheme that is well-controlled by observing the so-called truncation weight. We verified convergence in this parameter and confirmed that our main findings are not affected by the compression. The final time evolution was computed for a duration of $T=1000/J$ using a maximal truncation error of $\varepsilon=10^{-12}$ and a time step $\Delta t=0.1/J$. The steady state expectation values of the required observables are extracted by calculating the average over the last 2000 time steps and are shown in Fig.~\ref{fig:L12}.

Also for these larger systems one can nicely see a similar behaviour as described for $N=6$. As can be seen in Fig.~\ref{fig:L12}(a), the behaviour around the point $\Delta^\ast_{\pm 5}$ shows that only the winding numbers $m = \pm 5$ have an appreciable amplitude and the amplitudes of the other winding numbers rise slowly in its neighbourhood. This is compatible with the analytically expected rank-2 steady state decomposed of the two different winding numbers. The steepness of the rise of the amplitudes of additional winding numbers at the special points depends on the system size. In particular, with increasing system size the required value of $\Gamma$ in order to resolve the special point rises. This is accompanied by an exponential increase of the time-scales, such that it becomes very difficult to resolve the steady state in the Zeno limit at $\Delta^*$ for very large system sizes.

The approach of the Zeno limit can be clearly seen in the dependence on the value of $\Gamma$.  One finds that the expectation value of the boundary spins  collapses already for relatively low values of $\Gamma$ and becomes locked to the expected value of the dissipation-free subspace around the value of $\Gamma \approx 100J$ (not shown). This validates the interpretation that in the large $\Gamma$ limit the system is close to a ring in which the winding numbers can be associated with topological invariants. Further, as shown in Fig.~\ref{fig:L12} (b) for the value $\Delta^*_{\pm 5}$, the amplitudes of the winding numbers rapidly approach the expected values for the  predicted helical state for increasing $\Gamma$, i.e.~all amplitudes become negligible except for the amplitudes for $m\pm5$ which remains finite.


In the following we justify analytically the appearance of the steady state of rank two occuring in the Zeno limit at the points $\Delta^*$. To this end, we expand the density matrix of the steady state in orders of $1/\Gamma$ as
$\rho(\Gamma)=\sum_{n=0}^\infty \rho^{(n)}\Gamma^{-n}$.
Inserting this ansatz into the Lindblad equation one can decompose the equation in different orders.
The zero order condition leads to the condition that the density matrix of the boundary spins lies in the dissipation free subspace, i.e.~ $\rho^{(0)}=\ket{\uparrow_x}_1\bra{\uparrow_x}_1  \otimes R_0\otimes  \ket{\uparrow_x}_N\bra{\uparrow_x}_N$, where $R_0$ is the still undetermined bulk part.

In the first  order of expansion (see supplemental), we obtain the condition
\begin{align}
  [R_0,H_{\textrm{eff}}] &=0
  \label{CondCommutativity}
\end{align}
Here $H_{\textrm{eff}}$ acts in the Hilbert space of the internal bulk sites $2,\ldots N-1$ only. It is given by a $XXZ$-Hamiltonian with
boundary fields
\begin{align}
  H_{\textrm{eff}} &= \sum_{i=2}^{N-2}h^{XXZ}_{i,i+1}(\Delta)+ \frac{J}{2} S_2^x+\frac{J}{2} S_{N-1}^x - \frac{J}{2} \Delta
  \label{Heff}\\
h^{XXZ}_{i,j}(\Delta)  &=J\left[ S^x_i S^x_j +  S^y_i S^y_j + \Delta (S^z_i S^z_j - \frac{1}{4} I) \right].
  \label{Defhxxz}
\end{align}
For anisotropies $\Delta^*_m$ the helix states $\ket{\pm m}$ (\ref{SHS}), restricted to the internal sites $2$ to $ N-1$, are eigenstates
of  $H_{\textrm{eff}}$ with eigenvalue 0, i.e.~$H_{\textrm{eff}}\ket{\pm m}|_{\textrm{bulk}}=0$.
Thus, the condition (\ref{CondCommutativity}) is fulfilled by the ansatz  $R_0=b \ket{s}\bra{s}+ (1-b) \ket{a}\bra{a}$
which has rank $2$.

In order to find the weight $b$, we investigate the compatibility conditions arising in the second order in $1/\Gamma$.
Among other conditions (see suppl.~material) we obtain
\begin{align}
  b &=\frac{(1+ |\eta|)^2}{2 + 2 |\eta|^2}
  \label{Res_b}\\
  \eta&=\braket{m}{-m}|_{bulk} = \prod_{j=1}^{N-2} \cos{\left(\frac{2\pi j m}{N-1}\right)} = 2^{2-N}
  \label{Res_eta}
\end{align}
where the last equality holds for $m$ and $N-1$ coprime.
The overlap $\eta$ vanishes exponentially with system size and the predicted rank-2 steady state has contributions of the two helical states $\ket{\pm m}\bra{\pm m}$.

Further conditions (see supplemental material) need to be fulfilled by the steady state,
such that the rank-2 state Eq.~(\ref{eq:ResultNESSrank2}) is not necessarily the steady state. Considering our numerical findings (up to $N=13$), we come to the conjecture that the state Eq.~(\ref{eq:ResultNESSrank2})  is the true steady state at the fine tuned anisotropy $\Delta^*_{\pm m}$  in the Zeno limit, whenever $N-1$ and $m$ are coprime.

One very interesting open question which remains is what happens to these findings in the thermodynamic limit. In this limit the fine tuned values of the anisotropy become dense and the states of different winding numbers become close. It would be interesting to see whether the rank-2 steady states remain stable solutions and how a crossing between the different states can take place.


To summarize, we have found that helical states can be the steady states of a $XXZ$ model of finite size which is coupled at its boundaries to dissipation. We see that in this case the helical states are not protected by gaps in the Lindblad spectrum and that the transition between helical states with different winding numbers goes via highly mixed states. This opens the question whether other  examples exist of topologically interesting state in dissipatively driven systems which are not protected by a gap in the Lindbladian.

{\it Acknowledgments:} We thank S.~Diehl, M.~Fleischhauer and C. Presilla for fruitful discussions and A.~Sheikhan, C.~Halati for technical support. We acknowledge funding from the Deutsche Forschungsgemeinschaft (DFG, German Research Foundation) under project number KO 4771/3-1 (SCHU827/9-1), KL645/20-1 and
project number 277625399 - TRR 185 project B3 and project number 277146847 - project C05 and under Germany's Excellence Strategy – Cluster of Excellence Matter and Light for Quantum Computing (ML4Q) EXC 2004/1 – 390534769 and the European Research Council (ERC) under the Horizon 2020 research and innovation programme, grant agreement No. 648166 (Phonton) and 694544 (OMNES).

\bibliography{sessink_spinhelix_prl}

\clearpage

\setcounter{equation}{0}
\setcounter{page}{1}
\setcounter{figure}{0}

\onecolumngrid
\begin{center}
\Large
Supplemental material \\\vspace{.3cm} Transition between dissipatively stablized helical states
\end{center}
\vspace{1cm}
\twocolumngrid

\subsection{Perturbative argument for the rank-$2$ Zeno steady state}

Here we present details on the analytical justification  for a winding numbers $m$ around the $z$-axis. Our argumentation is based on a
perturbative expansion of the master equation in $1/\Gamma$. We make the ansatz for the steady state $\rho(\Gamma)=\sum_{k=0}^\infty \rho^{(k)}\Gamma^{-k}$.

Inserting the expansion of the steady state into the time-independent master equation $-i[H,\rho]+\Ga D[\rho]=0$,
and comparing the orders of $\Gamma^{1/k}$, we obtain recurrence relations for $k\geq 0$ given by
\begin{align}
D[\rho^{(0)}] &= 0\label{LME0}\\
D[\rho^{(k+1)}]-i [H,\rho^{(k)}]&= 0\label{LME2}
\end{align}
Taking the trace over the boundary sites $1$ and $N$, these relations lead to the requirement \cite{PopkovLivi2015}
\begin{align}
Tr_{1,N}[H,\rho^{(k)}] &=0\label{CondTr[H,rhok]}.
\end{align}

In the following we discuss how we can obtain the proposed rank-2 state in Eq.~(6) in the main text
from these relations.

The zeroth-order Eq.~(\ref{LME0}) only gives information at the boundary sites and is satisfied by the ansatz $\rho^{(0)}=\psi_0^1 \otimes R_0 \otimes \psi_0^N$ and $\psi_0^{1(N)} =\ket{\uparrow_x}_{1(N)}\bra{\uparrow_x}_{1(N)}$.

To obtain information about the bulk part of $R_0$, we need to consider the higher order relations. To obtain information from these, it is convenient to decompose the Hamiltonian as an operator acting
in the tensor product space ${\cal H}_0 \otimes {\cal H}_1$, where ${\cal H}_0$ is a Hilbert space of the two  boundary spins $1,N$,
and $ {\cal H}_1$ is the Hilbert space of the remaining  bulk spins $2,\ldots , N-1$. We introduce an orthonormal basis $e^{0},e^{1},e^{2},e^{3}$ in ${\cal H}_0$ by
\begin{align}
&e^{0}=\ket{\uparrow_x}_1 \otimes \ket{\uparrow_x}_N,\nonumber\\
&e^{1}=\ket{\downarrow_x}_1 \otimes \ket{\uparrow_x}_N,\nonumber\\
&e^{2}=\ket{\uparrow_x}_1 \otimes \ket{\downarrow_x}_N,\nonumber\\
&e^{3}=\ket{\downarrow_x}_1 \otimes \ket{\downarrow_x}_N.
\end{align}

The Hamiltonian with respect to this basis becomes
\begin{align}
 &H=\sum_{i,k} H_{i,k} \otimes  \ket{e^i}\bra{e^k}
  \label{DefSpectralDecomposition}\\
  &H_{i,k}= \bra{e^i}H\ket{e^k} 
  \label{Defgik}
\end{align}
One can show that the matrix elements between the zeroth and third state vanish, i.e.~ $$
  H_{0,3}=H_{3,0}=0.
$$
We introduce $g_{0,0} \equiv H_{\textrm{eff}}$ which is given by Eq.~(8)
in the main text. The commutator in Eq.~(\ref{LME2}) for $k=0$ can be rewritten using this decomposition as
\begin{align}
&[H,\rho^{(0)}]= \sum_{k=1,2} \left( H_{k,0} R_0 \ket{e^k} \bra{e^0} -  R_0 H_{0,k} \ket{e^0} \bra{e^k} \right) \nonumber\\
& + [H_{\textrm{eff}}, R_0]\otimes \ket{e^0} \bra{e^0} .
\end{align}

Using this representation and taking the trace over the boundary sites the condition simplifies to
\begin{align}
  [R_0,H_{\textrm{eff}}] &=0, \label{CondCommutativity-1}
\end{align}
which is given in Eq.~(7)
in the main text.
 The condition can be fulfilled if we assume the form
\begin{align}
  R_0 &= \sum_\al \nu_\al \ket {\al} \bra{\al}.
  \label{NESS-0}
\end{align}
Here $\ket{\alpha}$ are eigenvectors of $H_{\textrm{eff}}$ and $\nu_\alpha$ are some real valued, non-negative coefficients. They fulfill the condition $\sum _\alpha \nu_\alpha =1$ to give $Tr[\rho^{(0)}]=1$.
There exist some subtle issues connected to possible degeneracies of $H_{\textrm{eff}}$. These in particular can lead to the existence of steady states with higher ranks which goes beyond the scope of the current paper \cite{PopkovPresilla2019}).

Further, we can use the representation of the commutator in order to obtain information about $\rho^{(1)}$ from  Eq (\ref{LME2})
using the relations
\begin{align}
&D[\ket{e^k} \bra{e^0} ]= -\frac{1}{2} \ket{e^k} \bra{e^0}, \hspace{0.5cm} k=1,2 \\
&D[\ket{e^0} \bra{e^k} ]= -\frac{1}{2} \ket{e^0} \bra{e^k}  \hspace{0.5cm} k=1,2.
\end{align}
We obtain
\begin{align}
\rho^{(1)} &&= -2 i \sum_{k=1,2} \left( H_{k,0} R_0 \ket{e^k} \bra{e^0} -  R_0 H_{0,k} \ket{e^0} \bra{e^k} \right)\nonumber\\
&&+M_1 \otimes \ket{e^0} \bra{e^0},
\end{align}
where $M_1 \otimes \ket{e^0} \bra{e^0}$ is an arbitrary element from the kernel of the dissipator $D$ to be
determined by higher orders of the recurrence relations.
Inserting the above into Eq.~(\ref{CondTr[H,rhok]}) for $k=1$, and again using Eq.~(\ref{DefSpectralDecomposition}),
we obtain after some algebra
\begin{align}
&Q= \frac{i}{2}Tr_{1,N} [H,\rho^{(1)}]=\nonumber\\
&\sum_{k=1}^2 \left(H_{0,k} H_{k,0} R_0 + R_0 H_{0,k} H_{k,0} -2  H_{k,0} R_0  H_{0,k} \right)\nonumber\\
& + \frac{i}{2} [H_{\textrm{eff}},M_1]=0\nonumber
\end{align}
Finally, noting $H_{0,k}=H_{k,0}^\dagger$ (see also \cite{PopkovSchmidt2017b}
for details), and writing down the matrix elements $\bra{\al} Q \ket{\al}=0$
 we obtain
after some straightforward algebra for any value of $\alpha$,
\begin{align}
 & \sum_{\be\neq \al} w_{\al \be}\nu_\be = \nu_\al \sum_{\be\neq \al} w_{\al \be},
  \label{MarkovProcessEquation}\\
 &w_{\al \be}= |\bra{\be} H_{1,0} \ket{\al}|^2+  |\bra{\be} H_{2,0} \ket{\al}|^2.
\end{align}
In Eq.~(\ref{MarkovProcessEquation})  we recognize the steady state equation of a Markov process  with $w_{\al \be}$ being the rate of the transition from the state $\al$ to state  $\be$.  The explicit form of $H_{1,0}$, $  H_{2,0}$ can be calculated from Eq.~(\ref{Defgik}) (see e.g.~\cite{PopkovSchmidt2017}) and is given by
\begin{align}
  H_{1,0} &= \frac{J}{2}( S_{2}^y+ i S_{2}^z) \\
  H_{2,0} &= \frac{J}{2}( S_{N-1}^y+ i S_{N-1}^z).
  \label{ResW}
\end{align}
Note, that the index of the spin operators denotes the sites to which the operator is applied.
The Perron-Frobenius theorem guarantees an existence of a unique solution of Eq.~(\ref{MarkovProcessEquation}) with nonnegative entries, which sum up to $1$.
The quantities $\nu_\al$,
thus, have the double meaning of the eigenvalues of Eq.~(\ref{NESS-0}) in the original quantum Markov process and of steady-state probabilities of configurations in a classical Markov process with rates $w_{\al \be}$ associated to it, see also \cite{PopkovSchuetz2018}.

Now, the rank-$2$ state assumption, in terms of the associated Markov process Eq.~(\ref{MarkovProcessEquation})
means that the two states $0,1$ form a closed set, with weights $b,1-b$ which is a generalization of an absorbing state.
The closed set property is $ w_{0, \be}=w_{1,\be}=0$ for all $\be>1$.
We have checked numerically that the closed set property is satisfied for our setup for all $N \leq 13$, when $N - 1$ is a prime number \cite{EssinkMSc}.
Thus, the equation (\ref{MarkovProcessEquation})
for $\al=0,1$ becomes a closed equation for $b$, i.e.~
\begin{align}
  b \ w_{01} &=  (1-b) w_{10},\label{MPfor01}
 \end{align}
 where
\begin{align}
  w_{\al \be} &=|\bra{\be} H_{1,0} \ket{\al}|^2 + |\bra{\be} H_{2,0} \ket{\al}|^2 \label{MarkovProcessRates01}\\
\end{align}
from which we obtain the weights $b$.

\subsection{Extraction of the long time values from time-dependent matrix product state results}
In this section we outline the approach for extracting the long time values presented in the manuscript. As we described in the main part of the manuscript one of the difficulty is the slow relaxation of the states towards the steady states for larger values of $\Gamma$ due to the Zeno effect. In order to address this challenge and to be able to access steady state expectation values in the regime of very strong dissipation, we first evolve a state initially prepared in the N\'eel state
\begin{equation}
\rho(t=0) = \vert \psi_{\mathrm{Neel}} \rangle \langle  \psi_{\mathrm{Neel}}  \vert, \text{where   } \vert \psi_{\mathrm{Neel}} \rangle \equiv \vert \uparrow \downarrow \uparrow \downarrow\ldots \rangle,
\end{equation}
with a comparably low dissipative coupling $\Gamma=10J$ and large truncation weight $\varepsilon=10^{-10}$ up to a time $t=2200/J$ for the Hamiltonian parameters of interest. This parameters and the duration of this first time region is chosen such that a relatively fast change of the winding amplitudes is observed which then converges towards a steady states value in the following evolution.

\begin{figure}[t]
\vspace{0.3cm}
\includegraphics[width=0.99\columnwidth]{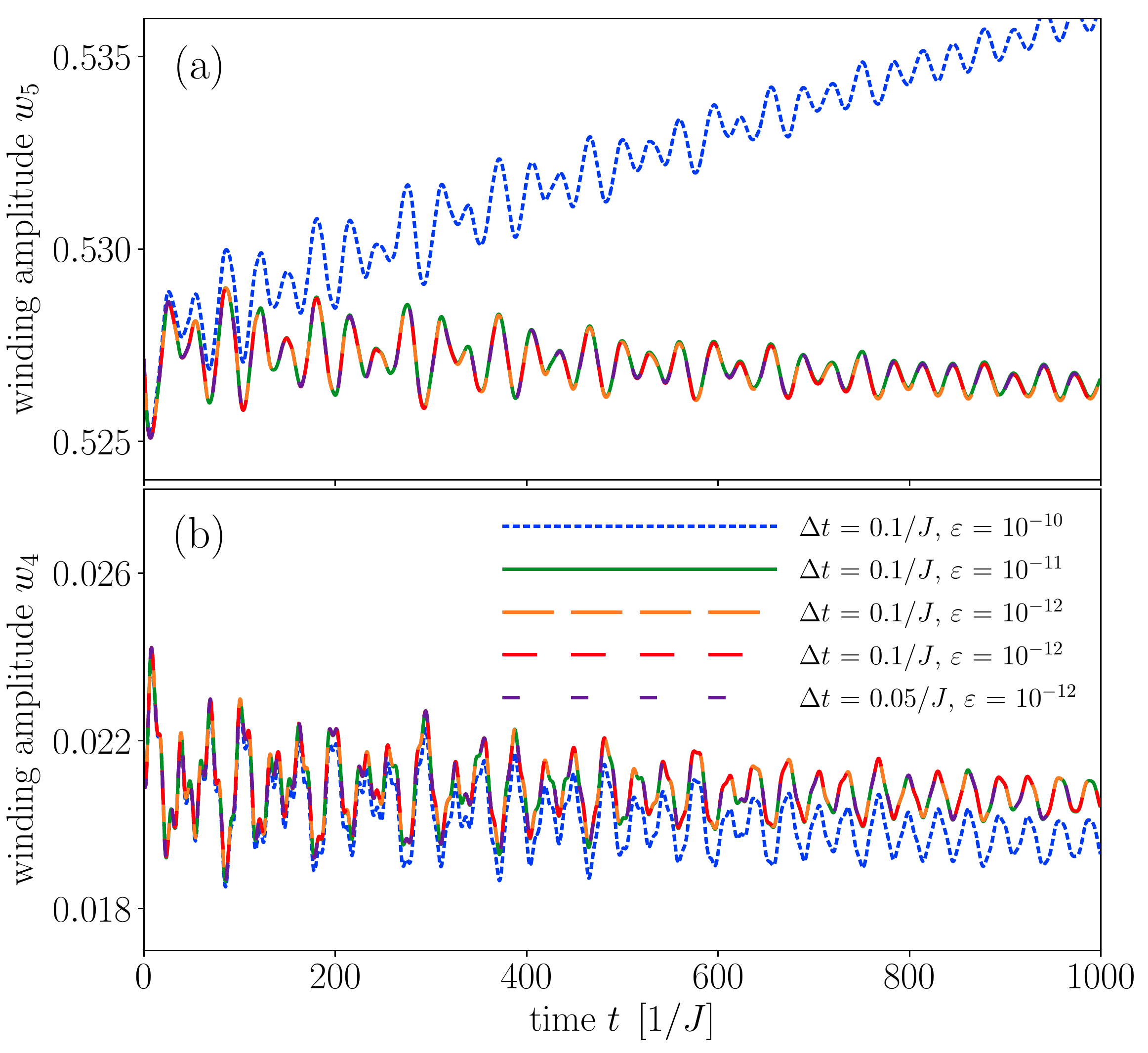}
\caption{(color online) Time-evolution of  the winding amplitudes (a) $w_5$ and (b) $w_4$. We present data for a chain of length $L=12$, anisotropy $\Delta = -0.87J$ and $\Gamma=50J$ and different values for truncation weight $\varepsilon$ and time step $\Delta t$. We find that all results collapse to the same curve and can be regarded as converged except for the case of $\varepsilon=10^{-10}$ and $\Delta t=0.1/J$.
  \label{fig:convergence} }
\end{figure}

As the initial evolution only aimed at providing a favourable initial state for the final evolution, the accuracy of the state representation was not of primary importance. In contrast we set up a subsequent evolution in order to converge to the real steady state using all correct parameters of the Lindblad evolution. This evolution is taken over a time window of $1000 /J$ and the convergence in terms of simulation parameters needs to be guaranteed in order to be able to make quantitative statements about the steady state. We present in Fig.~\ref{fig:convergence} of one exemplary parameter set with anisotropy $\Delta=-0.87J$ and dissipation strength $\Gamma=50J$ for different convergence parameters. As both, the truncation weight and the finite time step, introduce errors, verifying the agreement of the data for a certain combination of time step and truncation weight can be used as a measure of convergence. In the presented results, we see that the curve for $\varepsilon=10^{-10}$ and $\Delta t = 0.1J$ quickly deviates from the other curves, whereas the ones generated for smaller values of the truncation error and the time-step lie well on top of each other. We conclude that the choice of $\varepsilon=10^{-12}$ and $\Delta t=0.1J$ for the results in the main text is sufficiently well converged. Additionally, to the convergence in the numerical parameters, we see that the winding amplitudes are almost saturated at later time and only a very slow trend is still present between time $800/J$ and $1000/J$. We use this regime in order to extract the long time value presented in the main article.

\end{document}